\documentclass{acm_proc_article-sp}
\usepackage{amsmath}
\usepackage{algorithm}
\usepackage[noend]{algpseudocode}
\usepackage{url}
\usepackage{psfrag}
\usepackage{graphicx}

\begin{document}

\title{Asynchronous Parallel Stochastic Gradient Descent}
\subtitle{A Numeric Core for Scalable Distributed Machine Learning Algorithms}

\numberofauthors{1} 
\author{
\alignauthor
Janis Keuper and Franz-Josef Pfreundt\\\
       \affaddr{Fraunhofer ITWM}\\
       \affaddr{Competence Center High Performance Computing}\\
       \affaddr{Kaiserslautern, Germany}\\
       \email{\{janis.keuper | franz-josef.pfreundt\}@itwm.fhg.de}
}

\maketitle
\begin{abstract}
The implementation of a vast majority of machine learning (ML) algorithms boils down
to solving a numerical optimization problem. In this context, Stochastic 
Gradient Descent (SGD) methods have long proven to provide good results, both
in terms of convergence and accuracy. Recently, several parallelization approaches
have been proposed in order to scale SGD to solve very large ML problems. 
At their core, most of these approaches are following a MapReduce scheme.\\
This paper presents a novel parallel updating algorithm for SGD, which utilizes
the asynchronous single-sided communication paradigm.  
Compared to existing methods, Asynchronous Parallel Stochastic Gradient Descent
(ASGD) provides faster convergence, 
at linear scalability and stable accuracy.   
\end{abstract}

\section{Introduction}
The enduring success of Big Data applications, which typically includes 
the mining, analysis and inference of very large datasets, is leading to a change 
in paradigm for machine learning research objectives \cite{bottou2008tradeoffs}. 
With plenty data at hand, the traditional challenge of inferring generalizing 
models from small sets of available training samples moves out of focus. Instead,
the availability of resources like CPU time, memory size or network bandwidth 
has become the dominating limiting factor for large scale machine learning
algorithms.\\
In this context, algorithms which guarantee useful results even in the case
of an early termination are of special interest. With limited (CPU) time,
fast and stable convergence is of high practical value, especially when the 
computation can be stopped at any time and continued some time later when more
resources are available.\\   
\begin{figure}[t]
\includegraphics[width=\linewidth]{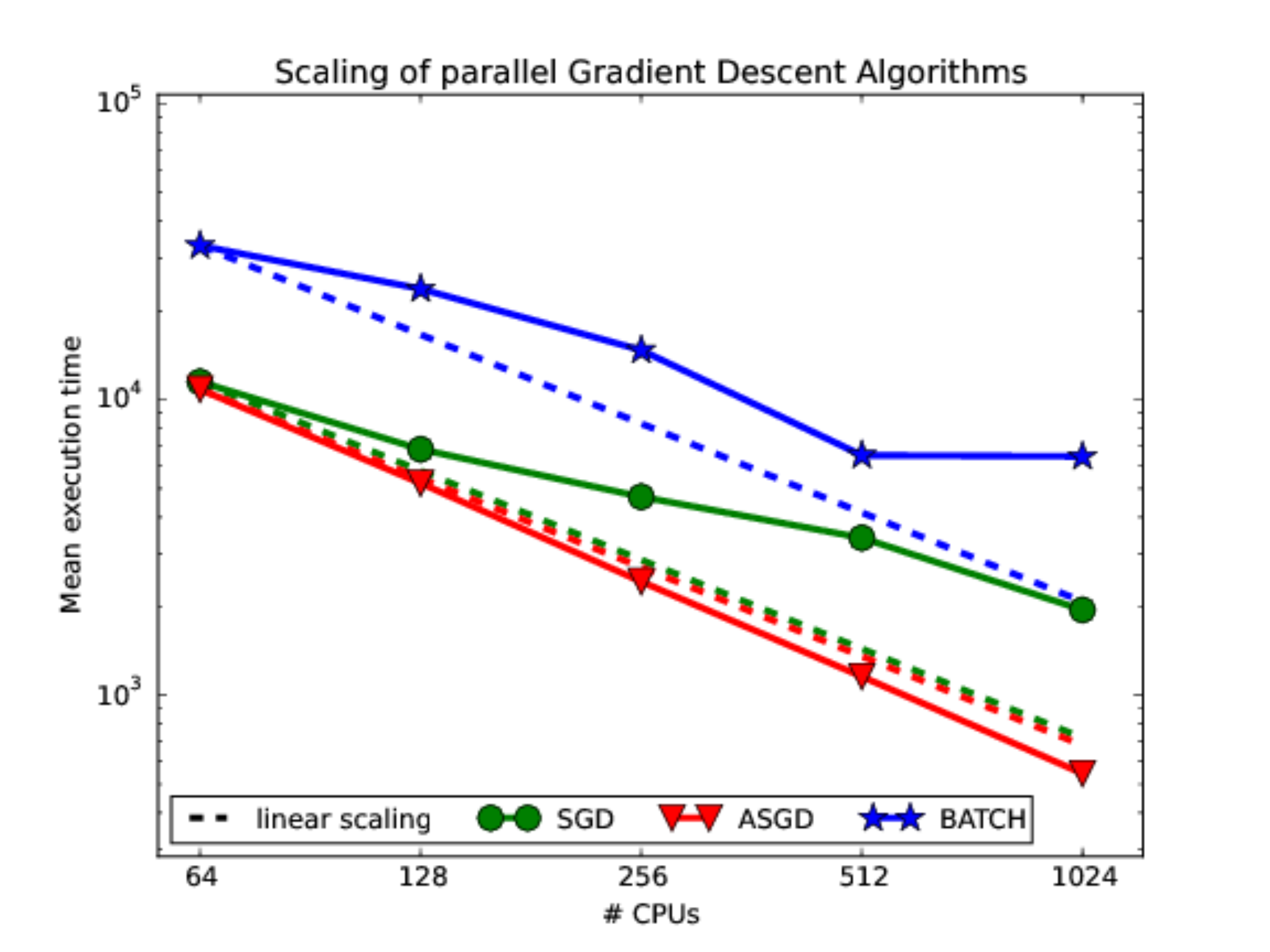}
\caption{
Evaluation of the scaling properties of different parallel gradient
descent algorithms for machine learning applications on distributed memory sytems. 
Results show a  
K-Means clustering
with k=10 on a  10-dimensional target space, represented by $\sim$1TB 
of training 
samples. Our novel ASGD method is not only the fastest algorithm in this test,
it also shows better than linear scaling performance. Outperforming the 
SGD parallelization by \cite{SGDsmola} and 
the MapReduce based BATCH \cite{chu2007map} optimization,
which both suffer from communication overheads.  
}
\end{figure}
\noindent Parallelization of machine learning (ML) methods has been a 
rising topic for some time (refer to \cite{datamining} for a comprehensive
overview). However, until the introduction of the MapReduce pattern, 
research was mainly focused on shared memory systems. This changed with
the presentation of a generic MapReduce strategy for ML algorithms in
\cite{chu2007map}, which showed that most of the existing ML techniques
could easily be transformed to fit the MapReduce scheme.\\   
After a short period of rather enthusiastic porting of algorithms to this 
framework, concerns started to grow if following the MapReduce ansatz 
truly provides a solid solution for large scale ML. It turns out, that 
MapReduce's easy parallelization comes at the cost of poor scalability
\cite{recht2011hogwild}. The main reason for this undesired behavior 
resides deep down within the numerical properties most machine learning 
algorithms have in common: an optimization problem. In this context, 
MapReduce works very well for the implementation of so called batch-solver
approaches, which were also used in the MapReduce framework of 
\cite{chu2007map}. However, 
batch-solvers have to run over the entire dataset to compute a single 
iteration step. Hence, their scalability with respect to the data size is
obviously poor.\\
Even long before parallelization had become a topic, most ML implementations 
avoided the known drawbacks of batch-solvers by usage of alternative online
optimization methods. Most notably, Stochastic Gradient Descent (SGD) methods 
have long  proven to provide good results for ML optimization problems \cite{recht2011hogwild}.
However, due to its inherent sequential nature, SGD is hard to parallelize and even harder 
to scale \cite{recht2011hogwild}.
Especially when communication latencies are causing dependency locks,
which is typical for parallelization tasks on distributed memory systems \cite{SGDsmola}.\\ 
The aim of this paper is to propose a novel, lock-free parallelization
method for the computation of stochastic gradient optimization for large scale 
machine learning algorithms on cluster 
environments.  

\subsection{Related Work} 
Recently, several approaches \cite{dean2012large}\cite{SGDsmola}\cite{sculley2010web}
\cite{recht2011hogwild}\cite{ngiam2011optimization} 
towards an effective parallelization
of the SGD optimization have been proposed. A detailed overview and in-depth
analysis of their application to machine learning can be found in \cite{SGDsmola}.\\
In this section, we focus on a brief discussion of four related publications, 
which provided the essentials for our approach:       
\begin{itemize} 
\item A theoretical framework for the analysis 
of SGD parallelization performance has been presented in \cite{SGDsmola}. 
The same paper also introduced a novel approach 
(called SimuParallelSGD), which 
avoids communication and any locking mechanisms up to a single and final 
MapReduce step. To the best of our knowledge, SimuParallelSGD 
is currently the best performing algorithm concerning cluster based 
parallelized SGD. A detailed discussion of this method is given in section 
\ref{sec_psgd}.
\item In \cite{sculley2010web}, a so-called mini-BATCH update scheme 
has been introduced. 
It was shown that replacing the strict online 
updating mechanism of SGD with small accumulations of gradient steps can 
significantly improve the convergence speed and robustness (also see
section \ref{sec_minisgd}).   
\item A widely noticed approach for a 
``lock-free'' parallelization of SGD on shared memory systems
has been introduced in \cite{recht2011hogwild}. The basic idea 
of this method is to explicitly ignore potential data races and to write
updates directly into the memory of other processes. Given a minimum level
of sparsity, they were able to show that possible data races will neither harm 
the convergence nor the accuracy of a parallel SGD. Even more, without
any locking overhead, \cite{recht2011hogwild} sets the current performance 
standard for shared memory systems.       
\item A distributed version of \cite{recht2011hogwild} has been presented in 
\cite{noel2014dogwild}, showing cross-host CPU to CPU and GPU to GPU gradient  
updates over Ethernet connections. 
\item In \cite{grunewald2013gaspi}, the concept of a  Partitioned 
Global Address Space programming framework (called GASPI) has been introduced. 
This provides an asynchronous, 
single-sided communication
and parallelization scheme for cluster environments (further details in section 
\ref{sec_gpi}). We build our asynchronous communication on the basis of this framework.     
\end{itemize}
\newpage
\subsection{Asynchronous SGD}\label{sec_ASGD_concept}
\begin{figure*}
\centering
\includegraphics[width=0.8\linewidth]{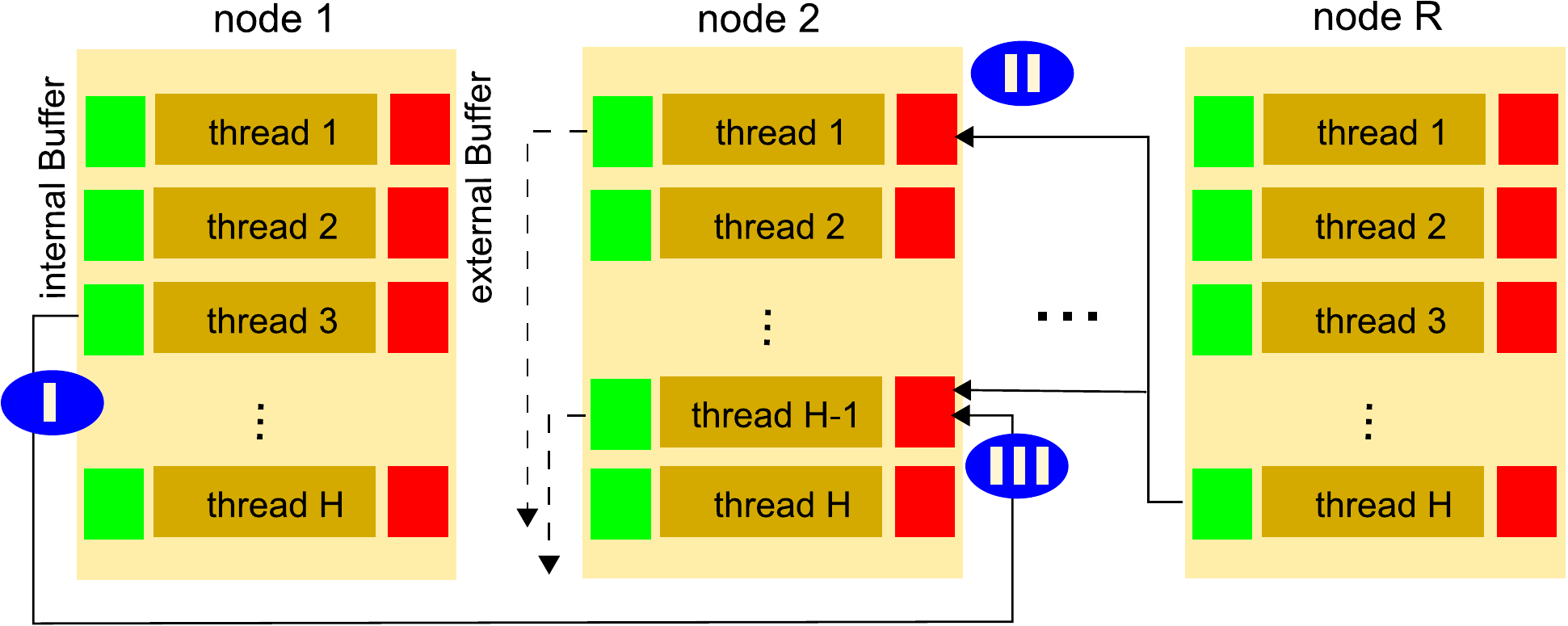}
\caption{Overview of the asynchronous update communication used in ASGD.
Given a cluster environment of $R$ nodes with $H$ threads each, the blue markers 
indicate different stages and scenarios of the communication mode. {\bf I:} Thread 
$3$ of node $1$ finished the computation of of its local  mini-batch update. 
The external buffer is empty. Hence it executes the update locally and sends
the resulting state to a few random recipients.
{\bf II:} Thread $1$ of node $2$ receives an update. When its local mini-batch update
is ready, it will use the external buffer to correct its local update and then
follow I.
{\bf III:} Shows a potential data race: two external updates might overlap
in the external buffer of thread $H-1$ of node $2$. Resolving
data races is discussed in section \ref{sec_race}.
\label{fig_ASGD}
} 
\end{figure*}

The basic idea of our proposed method is to port the ``lock-free'' shared memory approach from 
\cite{recht2011hogwild} to distributed memory systems. This is far from trivial,
mostly because communication latencies in such systems will inevitably cause 
expensive dependency locks if the communication is performed in common two-sided
protocols (such as MPI message passing or MapReduce). This is also the
motivation for SimuParallelSGD \cite{SGDsmola} to avoid communication
 during the optimization: locking costs are usually much higher than the information
gain induced by the communication.\\
We overcome this dilemma by the application of the asynchronous, single-sided 
communication model provided by \cite{grunewald2013gaspi}: individual processes
send mini-BATCH \cite{sculley2010web} updates completely uninformed 
of the recipients status whenever they are ready to do so. On the recipient 
side, available updates are included in the local computation as available. 
In this scheme, no process ever waits for any communication to be sent or 
received. Hence, communication is literally ``free'' (in terms of latency).\\
Of course, such a communication scheme will cause data races and race conditions: 
updates might  be (partially) overwritten before they are used or even might be 
contra productive because the sender state is way behind the state of the recipient.\\
We resolve these problems by two strategies: first, we obey 
the sparsity requirements introduced by \cite{recht2011hogwild}.
This can be 
achieved by sending only partial updates to a few random recipients. Second,
we introduce a Parzen-window function, selecting only those updates
for local descent which are likely to improve the local state.          
Figure \ref{fig_ASGD} gives a schematic overview of the ASGD algorithm's 
asynchronous communication scheme.
The remainder of this paper is organized as follows: first, we briefly review 
gradient descent methods in section \ref{sec_gd} and discuss further
aspects of the 
previously mentioned related approaches in more detail. Section \ref{sec_ac}
gives a quick overview of the asynchronous communication concept and its 
implementation. The actual details of the ASGD algorithm are introduced 
in section \ref{sec_asgd},
followed by a theoretical analysis and an extensive experimental evaluation in 
section \ref{sec_eval}.      

\section{Gradient Descent Optimization}\label{sec_gd}
From a strongly simplified perspective, machine learning tasks are usually about the 
inference of generalized models from a given dataset $X=\{x_0,\dots,x_m\}$ with $ x_i
\in\mathbb{R}^n$, 
which in case of supervised learning is also assigned with semantic labels 
$Y=\{y_0,\dots,y_m\}, y_i\in\mathbb{R}$.\\
During the learning process, the quality of a model is
evaluated by use of so-called loss-functions, which measure how well the current model 
represents the given data. We write $x_j(w)$ or $(x_j,y_j)(w)$ to indicate the 
loss of a data point for the current parameter set $w$ of the model function.
We will also refer to $w$ as the state of the model. The actual learning
is then the process of minimizing the loss over all samples.\\
This is usually implemented via a gradient descent over the partial derivative of
the loss function in the parameter space of $w$.   

\subsection{Batch Optimization}
The numerically easiest way to solve most gradient descent optimization
problems is the so-called batch optimization. A state $w_t$ at time $t$
is updated by the mean gradient generated by ALL samples of the available
dataset. Algorithm \ref{algo_BATCH} gives an overview of the BATCH optimization
scheme.  
\begin{algorithm}
\caption{BATCH optimization with
samples $X=\{x_0,\dots,x_m\}$,
iterations $T$, steps size
$\epsilon$ and
states $w$}
\label{algo_BATCH}
\begin{algorithmic}[1]
\ForAll{$t=0\dots T$ }
\State{\begin{bf}Init\end{bf} $w_{t+1}=0$ }
\State{\begin{bf}update\end{bf} $w_{t+1} = w_{t} - \epsilon\sum_{(X_j\in X)}\partial_wx_j(w_{t})$}
\State{$w_{t+1} = w_{t+1}/|X|$}
\EndFor
\end{algorithmic}
\end{algorithm}
A MapReduce parallelization for many BATCH optimized machine learning 
algorithms has been introduced by \cite{chu2007map}.

\subsection{Stochastic Gradient Descent}
In order to overcome the drawbacks of full batch optimization, many online 
updating methods have been proposed. One of the most prominent is SGD.
Although some properties of Stochastic Gradient Descent approaches might  
prevent their successful application to some optimization 
domains, they are well established in the machine learning community 
\cite{bottou2010large}.   
Following the notation in \cite{SGDsmola}, SGD can be formalized
in pseudo code as outlined in algorithm \ref{algo_SGD}. 
\begin{algorithm}
\caption{SGD with samples $X=\{x_0,\dots,x_m\}$, iterations $T$, steps size 
$\epsilon$ and states $w$}
\label{algo_SGD}
\begin{algorithmic}[1]
\Require{$\epsilon>0$}
\ForAll{$t=0\dots T$ } 
\State{\begin{bf}draw\end{bf} $j \in \{1\dots m\}$ uniformly at random} 
\State{\begin{bf}update\end{bf} $w_{t+1} \leftarrow w_{t} - \epsilon\partial_wx_j(w_{t})$}
\EndFor
\State{\Return $w_T$}
\end{algorithmic}
\end{algorithm}
The advantage in terms of computational cost with respect to the number of data 
samples is eminent: 
compared to batch updates of quadratic complexity,
SGD updates come at linearly growing iteration costs. At least for ML-applications, 
SGD error rates even outperform batch algorithms in many cases \cite{bottou2010large}.\\  

Since the actual update step in line $3$ of algorithm \ref{algo_SGD} plays a 
crucial role deriving our approach, we are simplifying the notation in this
step and denote the partial derivative of the loss-function for the remainder 
of this paper in terms of
an update step $\Delta$: 
\begin{equation}
\Delta_j(w_t) := \partial_wx_j(w_{t}). 
\end{equation}

\subsection{Parallel SGD\label{sec_psgd}}
The current ``state of the art'' approach towards a parallel SGD
algorithm for shared memory systems has been presented in \cite{SGDsmola}.
\begin{algorithm}
\caption{SimuParallelSGD 
with samples $X=\{x_0,\dots,x_m\}$, iterations $T$, steps size 
$\epsilon$, number of threads $n$ and states $w$}
\label{algo_SimuParallelSGD}
\begin{algorithmic}[1]
\Require{$\epsilon>0, n>1$}
\State{\begin{bf}define\end{bf} $H=\lfloor {m\over n}\rfloor$}
\State{randomly \begin{bf}partition\end{bf} $X$, giving $H$ samples to each node}
\ForAll{$i \in \{1,\dots,n\}$ \begin{bf}parallel\end{bf} }
\State{randomly {\bf shuffle} samples on node $i$}
\State{\begin{bf}init\end{bf} $w^i_{0}=0$}
\ForAll{$t=0\dots T$ }
\State{get the $t$th sample on the $i$th node and compute}
\State{\begin{bf}update\end{bf} $w^i_{t+1} \leftarrow w^i_{t} - \epsilon\Delta_t(w^i_t)$}
\EndFor 
\EndFor
\State{\begin{bf}aggregate\end{bf} $v = {1\over n}\sum_{i=1}^n w^i_{t}$}
\State{\Return $v$}
\end{algorithmic}
\end{algorithm}
The main objective in their ansatz is to avoid communication between working
threads, thus preventing dependency locks. After a coordinated initialization
step, all workers operate independently until convergence (or early termination).
The theoretical analysis in \cite{SGDsmola} unveiled the surprising fact that a single
aggregation of the distributed results after termination is sufficient in order to 
guarantee good convergence and error rates.\\
Given a learning rate (i.e. step size) $\epsilon$ and the number of threads $n$,
this formalizes as shown in algorithm \ref{algo_SimuParallelSGD}.

\subsection{Mini-Batch SGD}\label{sec_minisgd}
The mini-batch modification introduced by \cite{sculley2010web}
tries to unite the advantages of online SGD with the stability of BATCH
methods. It follows the SGD scheme, but instead of updating after each single
data sample, it aggregates several samples into a small batch. This mini
batch is then used to perform the online update.   
It can be implemented as shown in algorithm \ref{algo_mini}.
\begin{algorithm}
\caption{Mini-Batch SGD 
with samples $X=\{x_0,\dots,x_m\}$, iterations $T$, steps size
$\epsilon$, number of threads $n$ and mini-batch size $b$}
\label{algo_mini}
\begin{algorithmic}[1]
\Require{$\epsilon>0$}
\ForAll{$t=0\dots T$ }
\State{\begin{bf}draw\end{bf} mini-batch $M \leftarrow b$ samples from $X$}
\State{\begin{bf}Init\end{bf}${\bf\Delta}w_t=0$ }
\ForAll{$x\in M$ }
\State{\begin{bf}aggregate update\end{bf} ${\bf\Delta}w \leftarrow \partial_wx_j(w_{t})$}
\EndFor
\State{\begin{bf}update\end{bf} $w_{t+1} \leftarrow w_{t} - \epsilon{\bf\Delta}w_t$}

\EndFor
\State{\Return $w_T$}

\end{algorithmic}
\end{algorithm}

\section{Asynchronous Communication} \label{sec_ac} 
\begin{figure}[ht]
\includegraphics[width=\linewidth]{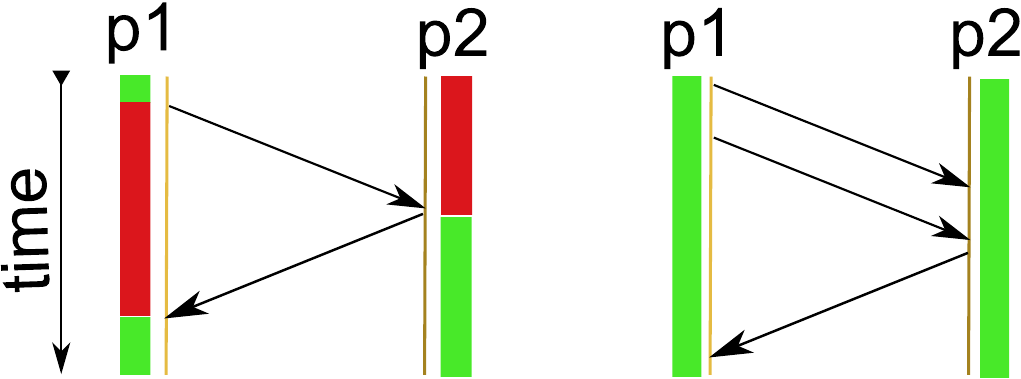}
\caption{Single-sided asynchronous communication model (right) compared to a typical 
synchronous model (left). The red areas indicate dependency locks of the 
processes $p1,p2$, waiting for data or acknowledgements. The asynchronous model
is lock-free, but comes at the price that processes never know if and when and 
in what order messages reach the receiver. Hence, a process can only be informed
about past states of a remote computation, never about the current status.   
\label{fig_async}
}
\end{figure}

Figure \ref{fig_async} shows the basic principle of the asynchronous 
communication model compared to the more commonly applied two-sided
synchronous message passing scheme.
An overview of the properties, theoretical implications and pitfalls of parallelization by 
asynchronous communication can be found in \cite{shan2012accelerating}. For the scope of this paper,
we rely on the fact that single-sided communication can be used to design lock-free
parallel algorithms. This can be achieved by design patterns propagating 
an early communication of data into work-queues of remote processes. Keeping
these busy at all times. If successful, communication virtually becomes ``free''
in terms of latency.\\
However, adopting sequential algorithms to fit such a pattern is far from trivial.
This is mostly because of the inevitable loss of information on the current state
of the sender and receiver.        

\subsection{GASPI Specification} \label{sec_gpi}
The Global Address Space Programming Interface (GASPI) \cite{grunewald2013gaspi}
 uses one-sided 
RDMA driven communication with remote completion to provide a scalable, flexible and
failure tolerant parallelization framework. GASPI favors an asynchronous communication 
model over two-sided bulk communication schemes. 
The open-source implementation GPI 2.0\footnote{Download available at http://www.gpi-site.com/gpi2/}
provides a C++ interface of the GASPI specification.\\
Benchmarks for various applications\footnote{Further benchmarks available at
http://www.gpi-site.com/gpi2/benchmarks/} show that 
the GASPI communication schemes can outperform MPI based implementations
\cite{grunewald2012bqcd} \cite{machado2011unbalanced} for many applications.

\section{The ASGD Algorithm} \label{sec_asgd} 
The concept of the ASGD algorithm, as described in section \ref{sec_ASGD_concept} 
and figure \ref{fig_ASGD} is formalized and implemented on the basis of the SGD 
parallelization presented in \cite{SGDsmola}. In fact, the asynchronous communication is 
just added to the existing approach. This is based on the assumption that communication 
(if performed correctly) can only improve the gradient descent - especially when it 
is ``free''. If the communication interval is set to infinity, ASGD will become 
SimuParallelSGD.    
\subsubsection*{Parameters}
ASGD takes several parameters which can have a strong influence on the convergence 
speed and quality (see experiments for details on the impact):\\
$\bf T$ defines the size of the data partition for each thread,
$\bf \epsilon$ sets the gradient step size (which needs to be fixed following
the theoretic constraints shown in \cite{SGDsmola}),
$\bf b$ sets the size of the  mini-batch aggregation, and
$\bf I$ gives the number of SGD iterations for each thread. Practically, this also equals the number 
of data points touched by each thread.

\subsubsection*{Initialization} 
The initialization step is straight forward and analog to SimuParallelSGD \cite{SGDsmola}
: the data is split into working packages of size $T$ and distributed to the 
worker threads. A control thread generates initial, problem dependent values for $w_0$
and communicates $w_0$ to all workers. From that point on, all workers run 
independently, following the asynchronous communication scheme shown in figure
\ref{fig_ASGD}.\\
It should be noted, that $w_0$ also could be initialized with the 
preliminary results of a previously early terminated optimization run.      

\subsubsection*{Updating}
The online gradient descent update step is the key leverage point of the ASGD 
algorithm. The local state $w^i_t$ of thread $i$ at iteration $t$ is updated 
by an externally modified step $\overline{\Delta_t(w^i_{t+1})}$, which not
only depends on the local $\Delta_t(w^i_{t+1})$ but also on a possible
communicated state $w^j_{t'}$ from an unknown iteration $t'$ at some random thread $j$:   
\begin{equation}
\overline{\Delta_t(w^i_{t+1})}=w^i_t-{1\over 2}\left( w^i_t + w^j_{t'} \right) + \Delta_t(w^i_{t+1})
\label{eq_ASGD_1}
\end{equation}

For the usage of $N$ external buffers per thread, we generalize equation (\ref{eq_ASGD_1}) to:
\begin{equation}
\begin{tabular}{l}
$\overline{\Delta_t(w^i_{t+1})}=w^i_t-{1\over |N|+1}\left( \sum_{n=1}^N\left(w^n_{t'}\right) 
    + w^i_t  \right) + \Delta_t(w^i_{t+1})$,\\
\quad\\where $|N|:=\sum_{n=0}^N \lambda(w^n_{t'}), \quad \lambda(w^n_{t'})= \left\{
\begin{tabular}{ll}
  $1$ & if $\|w^n_{t'}\|_2>0$ \\
  $0$ & otherwise 
\end{tabular}
\right.$
\end{tabular}
\label{eq_ASGD}
\end{equation}
with $N$ incoming messages. Figure \ref{fig_ASGD_pWindow} gives a schematic overview 
of the update process.

\subsection{{Parzen-Window} Optimization}
\begin{figure*}
\includegraphics[width=\linewidth]{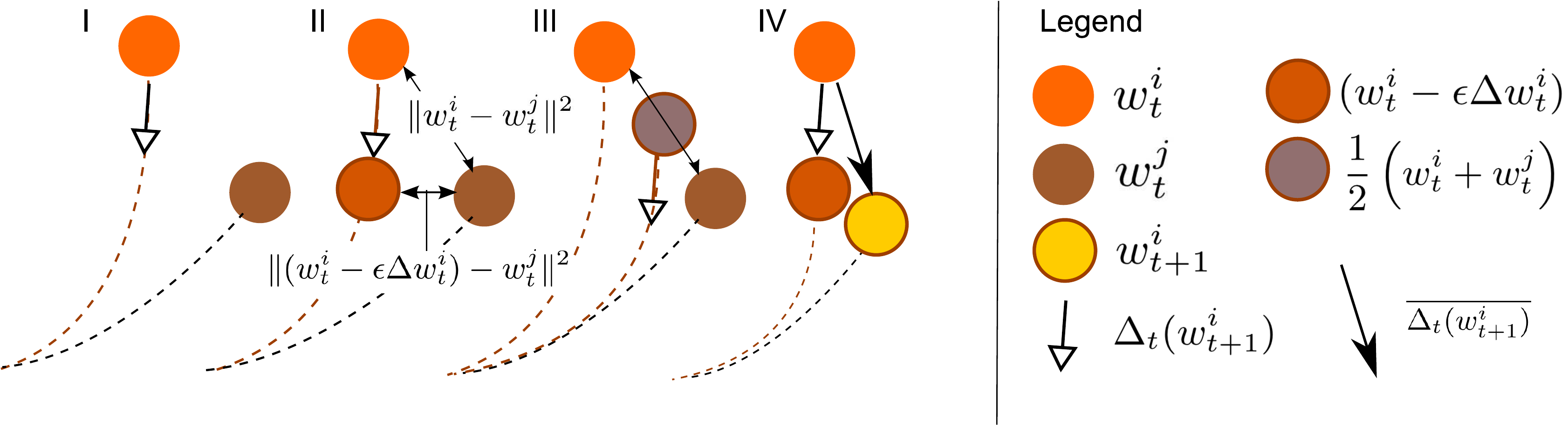}
\caption{ASGD updating. This figure visualizes the update algorithm 
of a process with state $w^i_t$, its local mini-batch update 
$\Delta_t(w^i_{t+1})$ and received external state  $w^j_t$ for a 
simplified 1-dimensional optimization problem. The dotted lines indicate 
a projection of the expected descent path to an (local) optimum. 
{\bf I:} Initial setting: ${\bf \Delta}_M(w^i_{t+1})$ is computed and $w^j_t$ 
is in the external
buffer.
{\bf II:} Parzen-window masking of $w^j_t$. Only if the condition of equation 
(\ref{eq_Parzen}) is
met, $w^j_t$ will contribute to the local update.
{\bf III:} Computing $\overline{{\bf\Delta}_M(w^i_{t+1})}$.
{\bf IV:} Updating $w^i_{t+1} \leftarrow w^i_t -\epsilon\overline{{\bf\Delta}_M
(w^i_{t+1})}$.   
\label{fig_ASGD_pWindow}
}
\end{figure*}

As discussed in \ref{sec_ASGD_concept} 
and shown in figure \ref{fig_ASGD}, the asynchronous communication scheme is prone 
to cause data races and other conditions during the update. Hence, we introduce a 
Parzen-window like function $\delta(i,j)$ to avoid ``bad'' update conditions. The data 
races are discussed in section \ref{sec_race}. 
\begin{equation}
\delta(i,j) := \left\{
\begin{tabular}{ll}
  $1$ & if $\|(w^i_t-\epsilon\Delta w^i_t)-w^j_{t'}\|^2 < \|w^i_t-w^j_{t'}\|^2$ \\
  $0$ & otherwise 
\end{tabular}
\right.,
\label{eq_Parzen}
\end{equation}
We consider an update to be ``bad'', if the external state $w^j_{t'}$ would direct
the update away from the projected solution, rather than towards it. Figure
\ref{fig_ASGD_pWindow} shows the evaluation of $\delta(i,j)$,
which is then plugged into the update functions of ASGD in order to exclude 
undesirable external states from the computation. Hence, equation (\ref{eq_ASGD_1}) 
turns into
\begin{equation}
\overline{\Delta_t(w^i_{t+1})}=\left[w^i_t-{1\over 2}\left( w^i_t + w^j_{t'} \right)\right]
\delta(i,j) + \Delta_t(w^i_{t+1})
\label{eq_ASGD_1_Parzen}
\end{equation}
and equation (\ref{eq_ASGD}) to
\begin{equation}
\begin{tabular}{rl}
$\overline{\Delta_t(w^i_{t+1})}=$&$w^i_t-1/\left(\sum_{n=1}^N\left(\delta(i,n)\right)+1\right)$\\
&$\cdot\left( \sum_{n=1}^N\left(\delta(i,n)w^n_{t'}\right) + w^i_t  \right)$\\
&$ + \Delta_t(w^i_{t+1})$
\end{tabular}
\label{eq_ASGD_Parzen}
\end{equation}

\subsubsection*{Computational Costs}
Obviously, the evaluation of $\delta(i,j)$ comes at some computational cost.
Since $\delta(i,j)$
has to be evaluated for each received message,
the ``free'' communication is actually not so free after all.
However, the costs are very low and can be reduced  
 to the computation of the distance between two states, which can be
achieved linearly in the dimensionality of the parameter-space of $w$ and the mini-batch size: 
$O({1\over b}|w|)$. Experiments in section \ref{sec_eval} show that the impact
of the communication costs are neglectable.\\
In practice, the communication frequency ${1\over b}$ is mostly constrained by the 
network bandwidth between the compute nodes, which is briefly discussed in section
\ref{sec_balance}.    

\subsection{Mini-Batch Extension}
We further alter the update of our ASGD by extending it with the mini-batch 
approach introduced
in section \ref{sec_minisgd}. The motivation for this is twofold: first, we would 
like to benefit from the advantages of mini-batch updates shown in 
\cite{sculley2010web}. Also, the sparse nature of the asynchronous communication 
forces us to accumulate updates anyway. Otherwise, the external states could only
affect single SGD iteration steps. Because the communication frequency is 
practically bound by the node interconnection bandwidth, the size of the 
mini-batch $b$ is used to control the impact of external states.\\    
We write ${\bf\Delta}_M$ in order to differentiate mini-batch steps
from single sample steps $\Delta_t$ of sample $x_t$:   
\begin{equation}
\overline{{\bf\Delta}_M(w^i_{t+1})}=\left[w^i_t-{1\over 2}\left( w^i_t + w^j_t \right)\right]
\delta(i,j) + 
{\bf\Delta}_M(w^i_{t+1})
\label{eq_ASGD_miniB}
\end{equation}
Note, that the step size $\epsilon$ is not independent of $b$ and should be 
adjusted accordingly.\\ 

\subsection{The final ASGD Update Algorithm\label{sec_final}}
Reassembling our extension into SGD, we yield the final ASGD algorithm.
With mini-batch size $b$, number of iterations $T$ 
and learning rate $\epsilon$ the update can be implemented like this:
\begin{algorithm}
\caption{ASGD $(X=\{x_0,\dots,x_m\},T,\epsilon,w_0,b)$}
\label{algo_ASGD}
\begin{algorithmic}[1]
\Require{$\epsilon>0, n>1$}
\State{\begin{bf}define\end{bf} $H=\lfloor {m\over n}\rfloor$}
\State{randomly \begin{bf}partition\end{bf} $X$, giving $H$ samples to each node}
\ForAll{$i \in \{1,\dots,n\}$ \begin{bf}parallel\end{bf} }
\State{randomly {\bf shuffle} samples on node $i$}
\State{\begin{bf}init\end{bf} $w^i_{0}=0$}
\ForAll{$t=0\dots T$ }
\State{\begin{bf}draw\end{bf} mini-batch $M \leftarrow b$ samples from $X$}
\State{\begin{bf}update\end{bf} $w^i_{t+1} \leftarrow w^i_{t} - \epsilon\overline{{\bf\Delta}_M(w^i_{t+1})}$}
\State{\begin{bf}send\end{bf} $w^i_{t+1}$ to random node $\neq i$} 
\EndFor
\EndFor
\State{\Return  $w^1_I$}
\end{algorithmic}

\end{algorithm}
At termination, all nodes $w^i, i \in \{1,\dots,n\}$ hold small local 
variations of the global
result. As shown in algorithm \ref{algo_ASGD}, one can simply return 
one of these local models (namely $w^1_I$) as global result. 
Alternatively, we could also aggregate the $w^i_I$ via map reduce.
Experiments in section \ref{sec_ex_agg} show that in most cases the first variant is 
sufficient and faster. 

\subsection{Data races and sparsity\label{sec_race}}
Potential data races during the asynchronous external update come in two forms: 
First, the complete negligence of an update state $w^j$ because it has been 
completely overwritten by a second state $w^h$. Since ASGD communication is 
de-facto optional, a lost message might slow down the convergence by a margin,
but is completely harmless otherwise. The second case is more complicated: a
partially overwritten message, i.e. $w^i$ reads an update from $w^j$ while this is overwritten
by the update from $w^h$.\\   
We address this data race issue based on the findings in \cite{recht2011hogwild}.
There, it has been shown that the error which is induced by such data races 
during an SGD update is linearly bound in the number of conflicting variables 
and tends to underestimate the gradient projection. \cite{recht2011hogwild}
also showed that for sparse problems, where the probability of conflicts is 
reduced, data race errors are negligible. For non sparse problems, 
\cite{recht2011hogwild} showed that sparsity can be induced by 
partial updating. We apply this approach to ASGD updates, leaving the 
choice of the partitioning to the application, e.g. for K-Means
we partition along the individual cluster centers of the states.
Additionally, the asynchronous communication model causes further sparsity
in time, as processes read external updates with shifted delays. This
further decreases the probability of races.               

\subsection{Communication load balancing\label{sec_balance}}
We  previously discussed that the choice of the communication frequency 
${1\over b}$ has a significant impact on the convergence speed.
Theoretically, more communication should be beneficial. However, due to
the limited bandwidth, the practical limit is expected to be far from $b=1$.\\ 
The choice of an optimal $b$ strongly depends on the data (in terms of dimensionality) 
and the computing environment:
interconnection bandwidth, number of nodes, CPUs per node, NUMA layout and 
so on. Hence, $b$ is a parameter which needs to be determined experimentally.\\
For most of the experiments shown in section
\ref{sec_eval}, we found $500\leq b \leq 2000$ to be quite stable.

\section{Experiments} \label{sec_eval}
We evaluate the performance of our proposed method in terms of convergence speed, 
scalability and error rates of the learning objective function using the 
K-Means Clustering algorithm. The motivation to choose this algorithm for
evaluation is twofold: First, K-Means is probably one of the simplest machine
learning algorithms known in the literature (refer to \cite{jain2010data} for a 
comprehensive overview). This leaves little room for algorithmic optimization 
other than the choice of the numerical optimization method. Second, it is also
one of the most popular\footnote{The original paper \cite{lloyd1982least} has been
cited several thousand times.} unsupervised learning algorithms with a wide 
range of applications and a large practical impact.        

\subsection{K-Means Clustering}
K-Means is an unsupervised learning algorithm, which tries to find the
underlying cluster structure of an unlabeled vectorized dataset.
Given a set of $m$ $n$-dimensional points $X=\{x_i\},i=1,\dots,m$, which is to
be clustered into a set of $k$ clusters, $w=\{w_k\},k=1,\dots,k$. The K-Means
algorithm finds a partition such that the squared error between the
empirical mean of a cluster and the points in the cluster is minimized.\\
It should be noted, that finding the global minimum of the squared error
over all $k$ clusters $E(w)$ is proven to be 
a NP-HARD problem \cite{jain2010data}. Hence, all optimization methods 
investigated in this paper are only approximations of an optimal solution.   
However, it has been shown \cite{Meila}, that K-Means finds local optima 
which are very likely to be in close proximity to the global minimum if the 
assumed structure of $k$ clusters is actually present in the given data.  

\subsubsection*{Gradient Descent Optimization}
Following the notation given in \cite{bottou1994convergence}, K-Means is 
formalized as minimization problem of the quantization error $E(w)$: 
\begin{equation}
E(w)=\sum_i{1\over 2}(x_i-w_{s_i(w)})^2,
\label{eq_kmeans}
\end{equation} 
where $w=\{w_k\}$ is the target set of $k$ prototypes for given $m$ examples 
$\{x_i\}$ and $s_i(w)$ returns the index of the closest prototype to the sample 
$x_i$.
The gradient descent of the quantization error $E(w)$ is then derived as 
$\Delta(w)={\partial E(w)\over \partial w}$. For the usage with the 
previously defined gradient descent algorithms, 
this can be reformulated to the following update
functions with step size $\epsilon$. 
Algorithms \ref{algo_BATCH} and \ref{algo_ASGD} use a batch update scheme.
Where the size $m' = m$ for the original BATCH algorithm and $m' << m$ for 
our ASGD:
\begin{equation}
\Delta(w_k)={1\over m'}\sum_i\left\{
\begin{tabular}{ll}
  $x_i-w_k$ & if $k=s_i(w)$ \\
  $0$ & otherwise 
\end{tabular}
\right.
\label{eq_km_batch}
\end{equation}

The SGD (algorithm \ref{algo_SimuParallelSGD}) uses an online update:
\begin{equation}
\Delta(w_k)=\left\{
\begin{tabular}{ll}
  $x_i-w_k$ & if $k=s_i(w)$ \\
  $0$ & otherwise 
\end{tabular}
\right.
\label{eq_km_online}
\end{equation}

\subsubsection*{Implementation}
We applied all three previously introduced gradient descent methods to 
K-Means clustering: the batch optimization with 
MapReduce
\cite{chu2007map} (algorithm \ref{algo_BATCH}), 
the parallel SGD \cite{SGDsmola} (algorithm \ref{algo_SimuParallelSGD})
 and our proposed ASGD (see algorithm \ref{algo_ASGD}) 
method. We used the C++ interface of GPI 2.0 \cite{grunewald2013gaspi} 
and the C++11
standard library threads for local parallelization.\\
To assure a fair comparison, all methods share the same data IO and distribution 
methods, as well as an optimized MapReduce method, which uses a tree 
structured communication model to avoid transmission bottlenecks.   

\subsection{Cluster Setup}
The experiments were conducted on a Linux cluster with a BeeGFS\footnote{see www.beegfs.com for details}
parallel file system.  
Each compute node is equipped
 with dual Intel Xeon E5-2670, totaling to 16 CPUs per node, 32 GB RAM and 
interconnected with FDR Infiniband.\\
If not noted otherwise, we used a standard of 64 nodes to compute the 
experimental results (which totals to 1024 CPUs). 

\subsection{Data\label{sec_data}}
We use two different types of datasets for the experimental evaluation
and comparison of the three investigated algorithms: a synthetically generated
collection of datasets and data from an image classification application.
\subsubsection*{Synthetic Data Sets}
The need to use synthetic datasets for evaluation arises from several rather 
profound reasons: (I) the optimal solution is usually unknown for real data,
(II) only a few very large datasets are publicly available, and, (III) we even 
need a collection of datasets with varying parameters such as dimensionality $n$,
size $m$ and number of clusters $k$ in order to evaluate the scalability.\\
The generation of the data follows a simple heuristic: given $n,m$ and $k$ we
randomly sample $k$ cluster centers and then randomly draw $m$ samples. Each
sample is randomly drawn from a distribution which is uniquely generated for 
the individual centers. Possible cluster overlaps are controlled by additional
minimum cluster distance and cluster variance parameters. The detailed properties 
of the datasets are given in the context of the experiments.     
\subsubsection*{Image Classification}  
Image classification is a common task in the field of computer vision. Roughly
speaking, the goal is to automatically detect and classify the content of
images into a given set of categories like persons, cars, airplanes, bikes,
furniture and so on. A common approach is to extract low level image features 
and then to generate a ``Codebook'' of universal image parts, the so-called 
Bag of Features \cite{nowak2006sampling}. Objects are then described 
as statistical model of these parts. The key step towards the generation of
the ``Codebook'' is a clustering of the image feature space.\\
In our case, large numbers of $d=128$ dimensional HOG features 
\cite{zhu2006fast} were extracted from a collection of images and clustered 
to form ``Codebooks'' with $k=100,\dots,1000$ entries.   
 
\subsection{Evaluation}
Due to the non-deterministic nature of stochastic methods and the fact that 
the investigated K-Means algorithms might get stuck in local minima, we  
apply a 10-fold evaluation of all experiments. If not noted otherwise, plots
show the mean results. Since the variance is usually magnitudes lower than
the plotted scale, we neglect the display of variance bars in the plots for the 
sake of readability. If needed, we report significant differences in the variance
statistics separately. To simply the notation, we will denote the SimuParallelSGD \cite{SGDsmola}
algorithm by SGD, the MapReduce baseline method \cite{chu2007map} by BATCH and
our algorithm by ASGD.   
For better comparability, we give the number of iterations $I$
as global sum over all samples that have been touched by the respective algorithm.
Hence, $I_{BATCH}:=T\cdot|X|$, $I_{SGD}:=T\cdot|CPUs|$ and 
$I_{ASGD}:=T\cdot b\cdot|CPUs|$.\\ 
Given runtimes are computed for optimization only, neglecting the initial data transfer
to the nodes, which is the same for all methods.       
Errors reported for the synthetic datasets are computed as follows: We use the
``ground-truth'' cluster centers from the data generation step to measure their
distance to the centers returned by the investigated algorithms. It is obvious
that this measure has no absolute value. It is only useful to compare the relative
differences in the convergence of the algorithms. Also, it can not be expected that
a method will be able to reach a zero error result. This is simply because 
there is no absolute truth for 
overlapping clusters which can be obtained from the generation process without
actually solving the exact NP-HARD clustering problem. Hence, the ``ground-truth''
is most likely also biased in some way.       


\begin{figure}[!t]
\includegraphics[width=\linewidth]{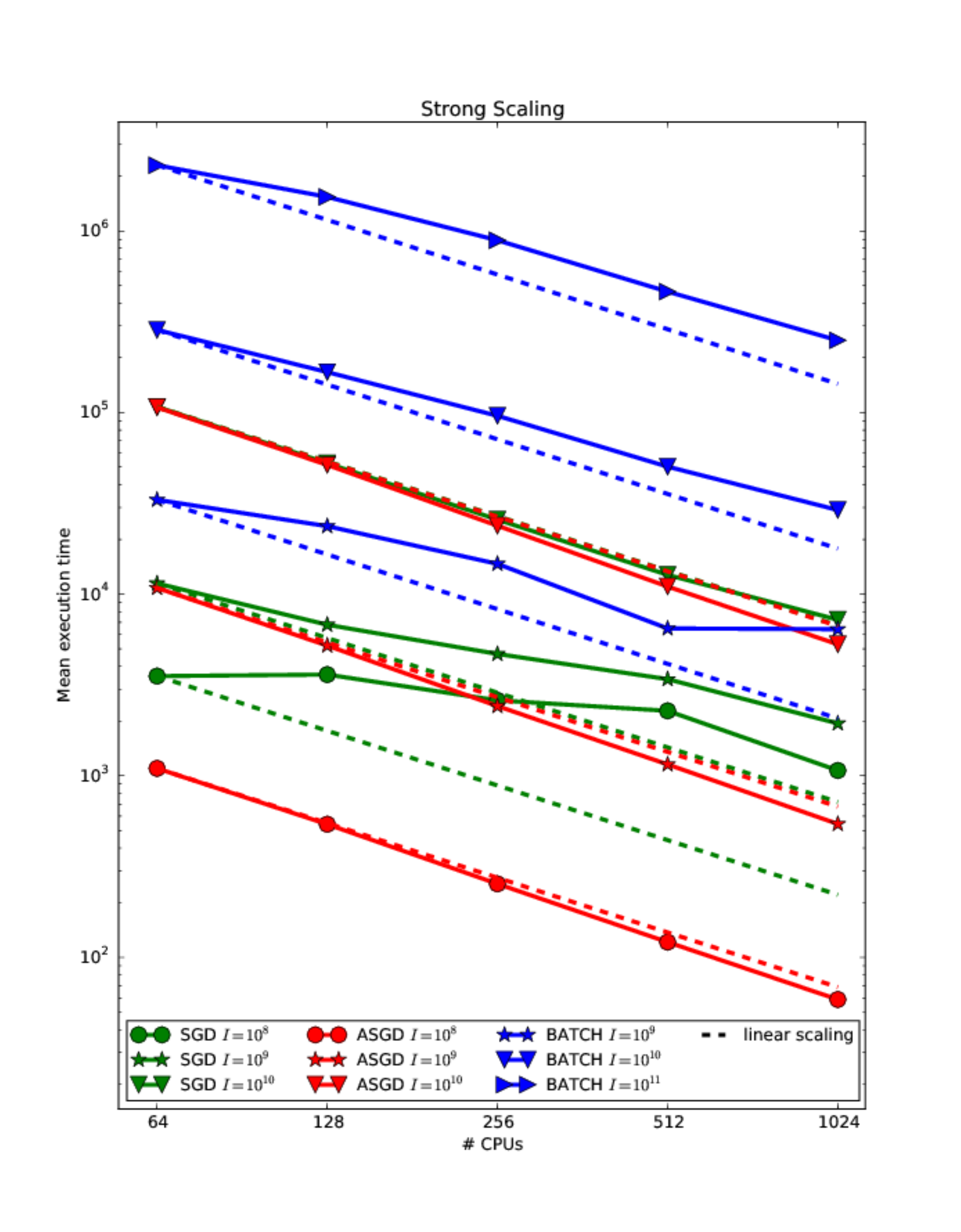}
\caption{Results of a strong scaling experiment on the synthetic dataset
with k=10, d=10 and $\sim$1TB data samples for different numbers of 
iterations $I$. The related error rates are shown in figure 
\ref{fig_scale_err}.
\label{fig_eval_scaling}}
\end{figure}
\subsection{Experimental Results}
\subsubsection*{Scaling\label{sec_ex_scale}}
We evaluate the runtime and scaling properties of our proposed algorithm 
in a series of experiments on synthetic and real datasets (see section 
\ref{sec_data}). First, we test a strong scaling scenario, where the 
size of the input data (in $k,d$ and number of samples) and the 
global number of iterations are constant for each experiment, 
while the number of CPUs is increased.
\begin{figure}[!ht]
\includegraphics[width=\linewidth]{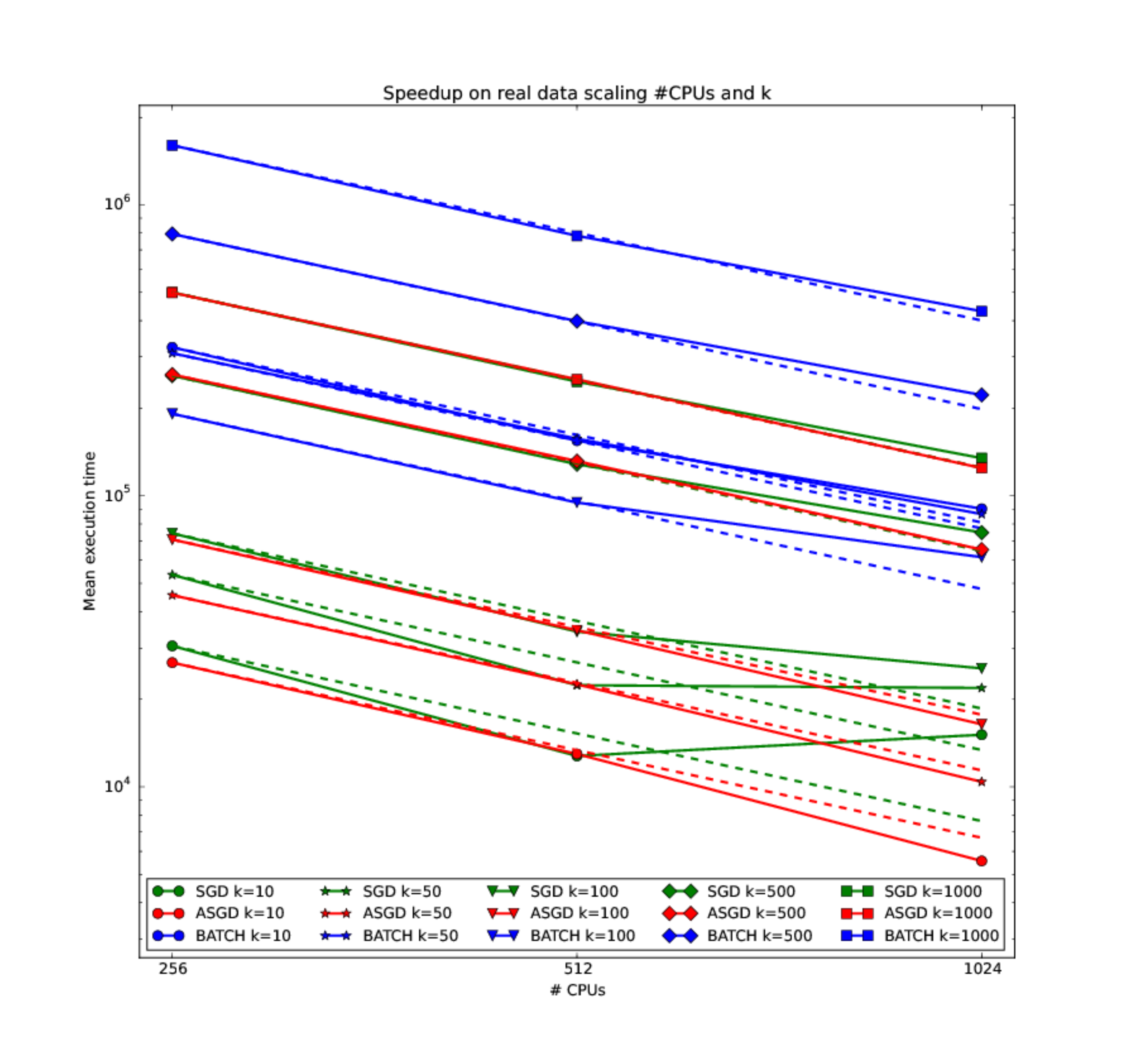}
\caption{Strong scaling of real data. Results for with $I=10^{10}$ 
and $k=10..1000$ on the image classification dataset.                                               
\label{fig_eval_real_scaling}}
\end{figure}
Independent of the number of iterations and CPUs, ASGD is always
the fastest method, both for synthetic (see figures \ref{fig_eval_real_scaling},\ref{fig_scale_err})
 and real data (figure \ref{fig_eval_real_scaling}).
Notably, it shows (slightly) better than linear
scaling properties. The SGD and BATCH methods suffer from a
communication overhead which drives them well beyond linear scaling
(which is projected by the dotted lines in the graphs). 
For SGD, this effect is dominant for
smaller numbers of iterations\footnote{Note: as shown in figure \ref{fig_scale_err},
a smaller number of iterations is actually sufficient to solve the given problem.} 
and softens proportionally with the increasing number of iterations. This is
due to the fact that the communication cost is independent of the number of 
iterations.\\
The second experiment investigates scaling in the number of target clusters
$k$, given constant $I,d$, number of CPUs and data size. Figure 
\ref{fig_eval_real_k} shows that all methods scale better than
$O(\log k)$. While ASGD is faster than the other methods, its scaling properties are
slightly worse. This is due to the fact that the necessary sparseness of
the asynchronous updates (see section \ref{sec_race}) is increasing with
$k$. 
\begin{figure}[!ht]
\includegraphics[width=\linewidth]{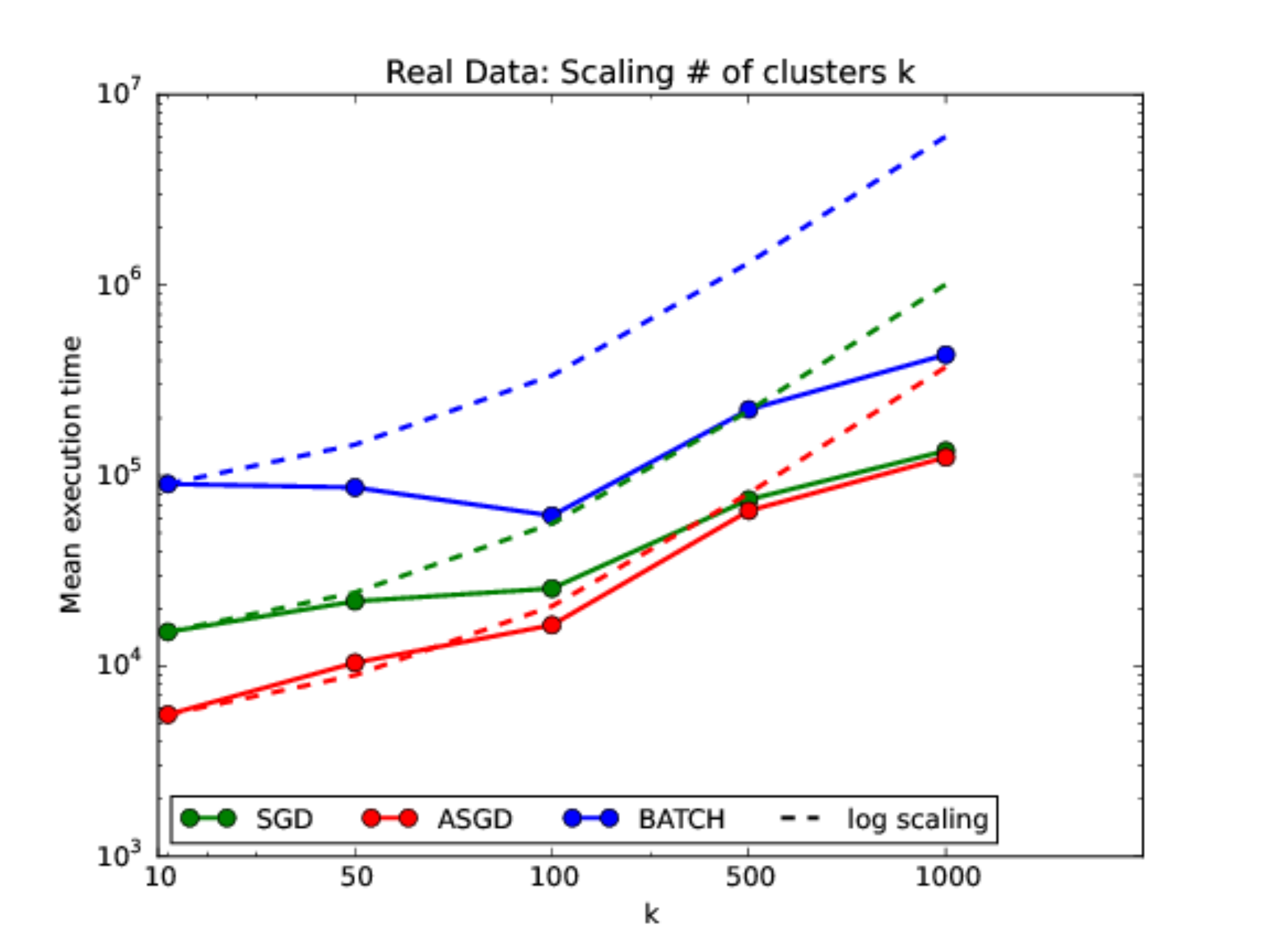}
\caption{Scaling the number of clusters $k$ on real data. Results
for the same experiment as in figure \ref{fig_eval_real_scaling}.
Note: here, the dotted lines project
a logarithmic scaling of the runtime in the number of clusters.
\label{fig_eval_real_k}}
\end{figure}
\newpage
\subsubsection*{Convergence Speed\label{sec_ex_con}}
\begin{figure}[!ht]
\includegraphics[width=\linewidth]{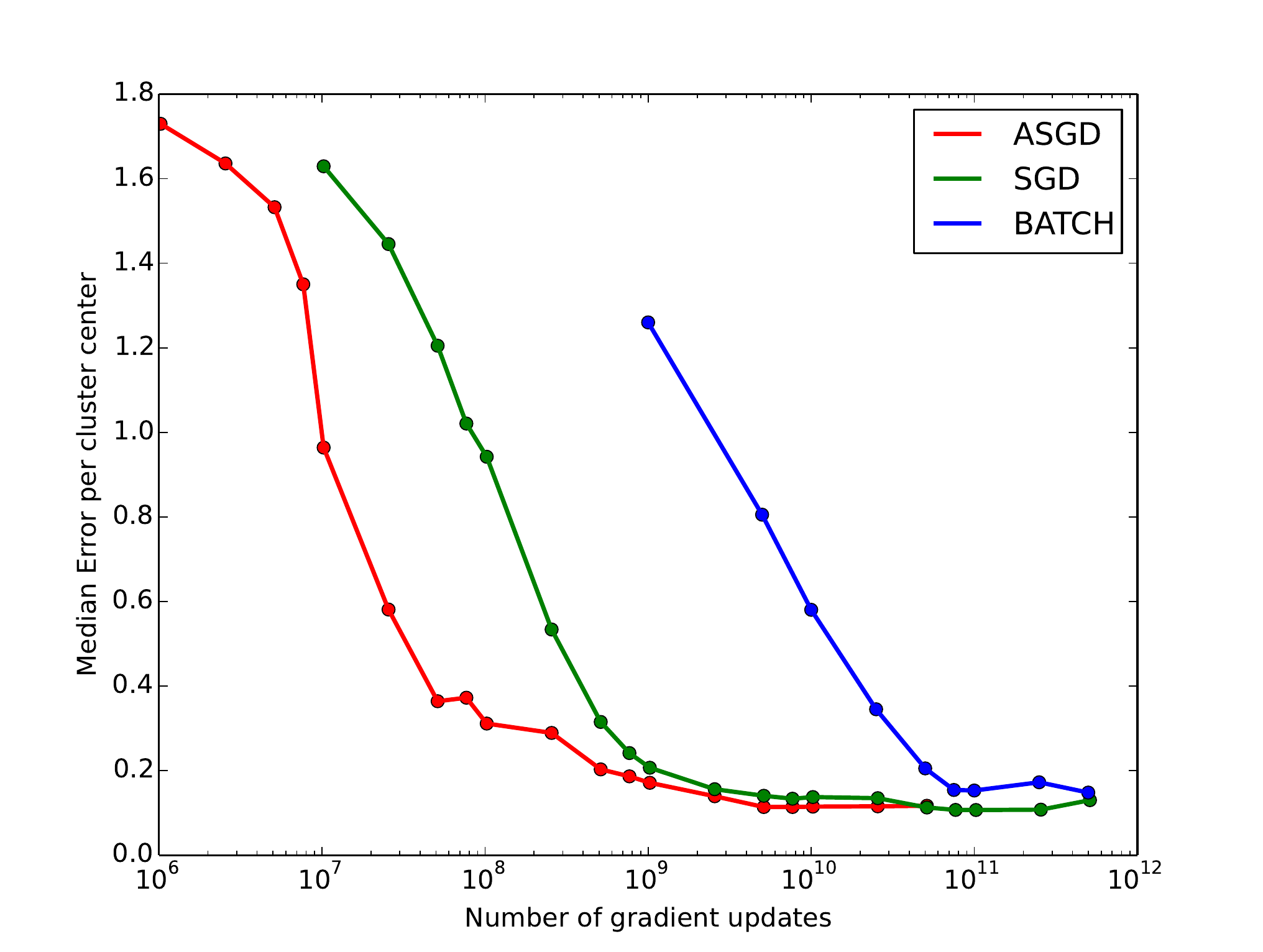}
\caption{Convergence speed of different gradient descent methods used to solve
K-Means clustering
with $k=100$ and $b=500$ on a $10$-dimensional target space parallelized  over $1024$ CPUs
on a cluster. Our novel ASGD method outperforms communication free SGD
\cite{SGDsmola} and
MapReduce
based BATCH \cite{chu2007map} optimization by the order of magnitudes.
\label{fig_eval_conv}
}
\end{figure}

Convergence (in terms of iterations and time) is an important factor in large 
scale machine learning, where the early termination properties of algorithms 
have a huge practical impact. Figure \ref{fig_eval_conv} shows the superior 
convergence properties of ASGD. While it finally converges to similar error 
rates, it reaches a fixed error rate with less iterations than SGD or BATCH.
As shown in figure \ref{fig_eval_conv}, this early convergence property
can result in speedups up to one order of magnitude.      
\newpage
\subsubsection*{Optimization Error after Convergence\label{sec_ex_err}} 
\begin{figure}[!ht]
\includegraphics[width=\linewidth]{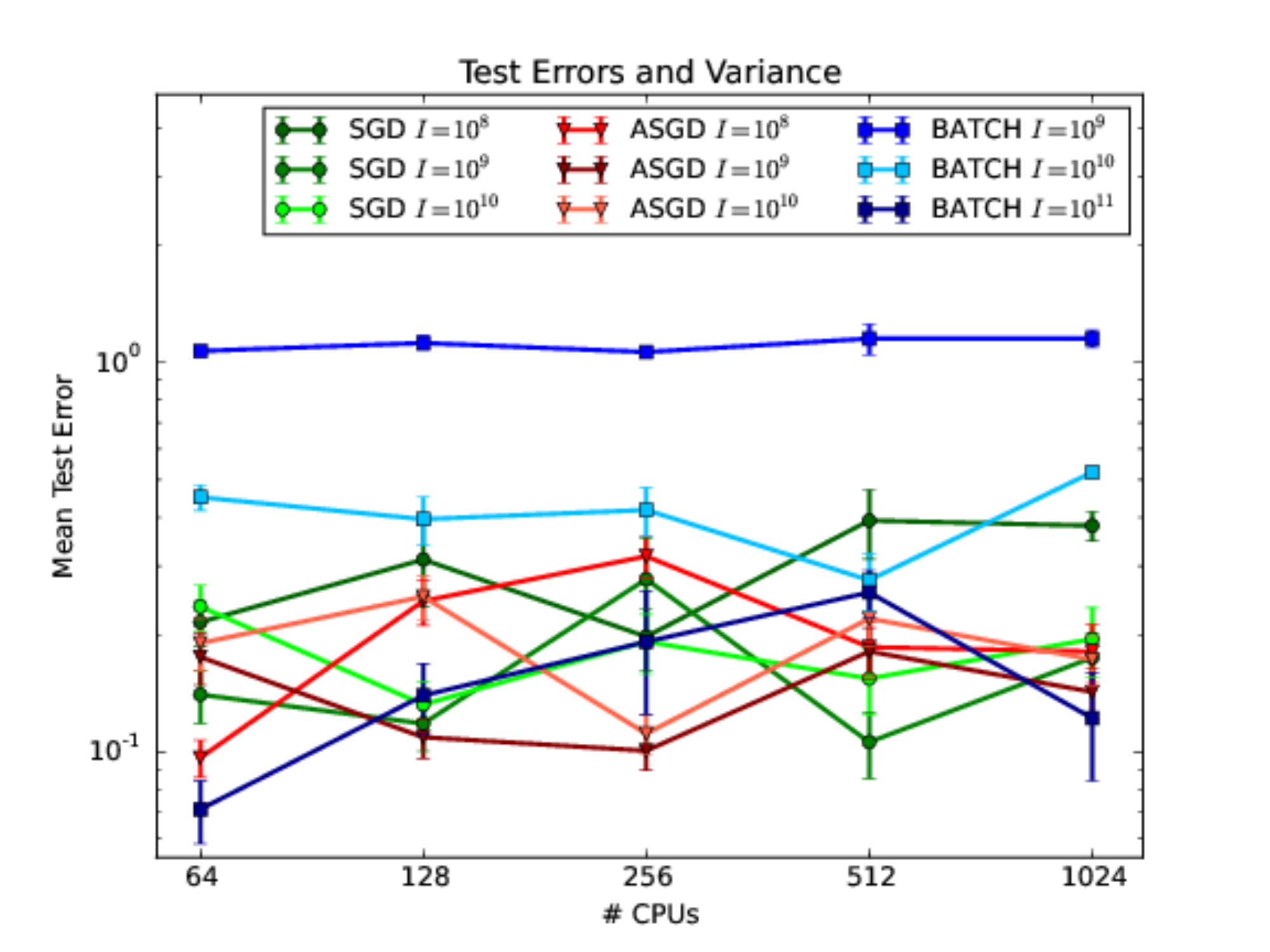}
\caption{Error rates and their variance of the strong scaling experiment 
on synthetic data shown in figure \ref{fig_eval_scaling}. A more detailed 
view of the variances is shown in figure \ref{fig_scale_var}.
\label{fig_scale_err}}
\end{figure}

\begin{figure}[!ht]
\includegraphics[width=\linewidth]{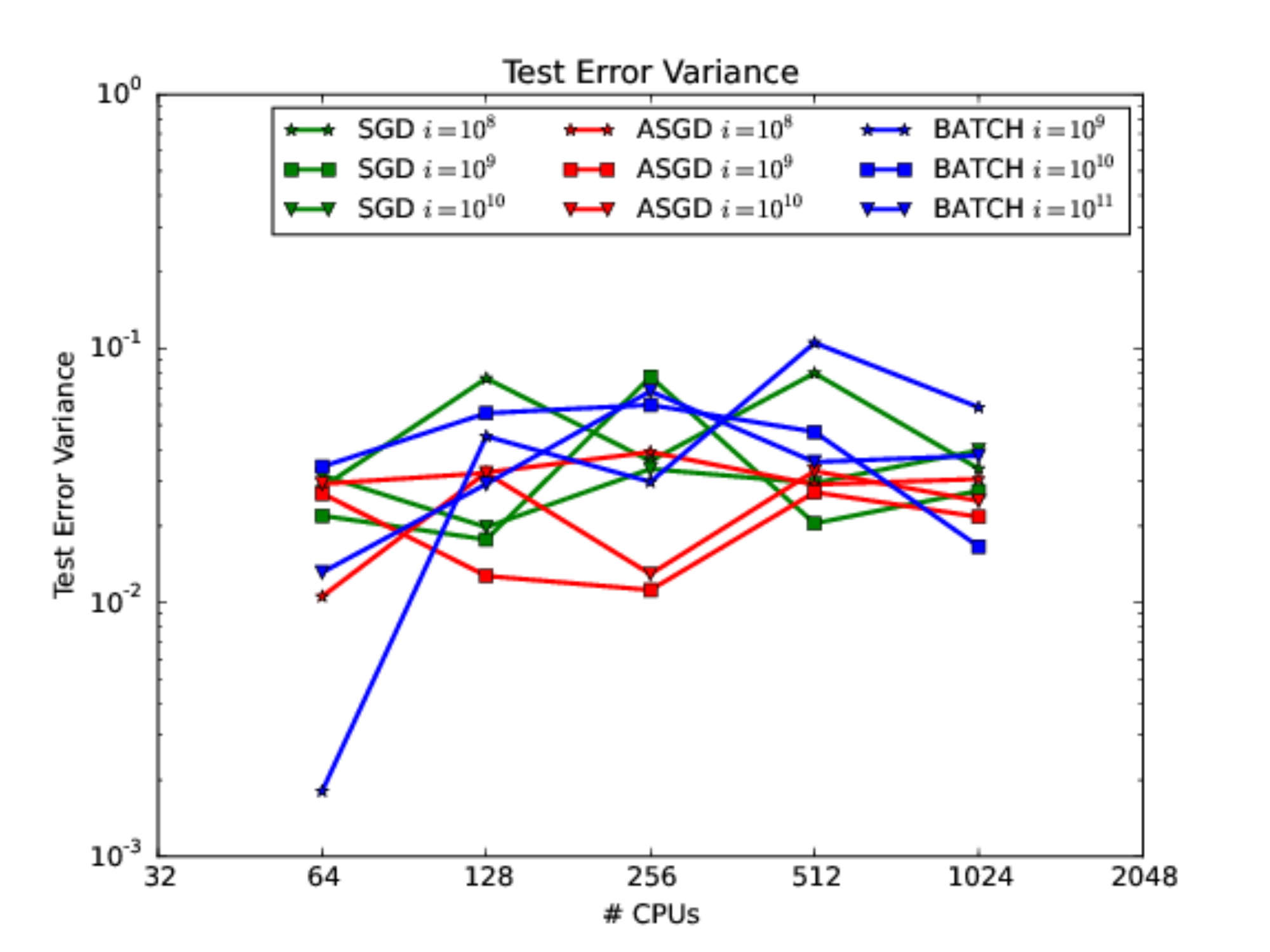}
\caption{Variance of the error rates of the strong scaling experiment 
on synthetic data shown in figure \ref{fig_eval_scaling}.
\label{fig_scale_var}}
\end{figure}

The optimization error after full convergence\footnote{Full convergence is 
here defined as the state where the error rate is not improving after 
several iterations.} for the strong scaling experiment (see section \ref{sec_ex_scale})
is shown in figures \ref{fig_scale_err} and \ref{fig_scale_var}. While ASGD 
outperforms BATCH, it has no significant difference in the mean error rates
compared to SGD. However, figure \ref{fig_scale_var} shows, that it 
tends to be more stable in terms of the variance of the non-deterministic
K-Means results.  

\subsubsection*{Communication Frequency\label{sec_ex_com}}
\begin{figure}[ht]
\includegraphics[width=\linewidth]{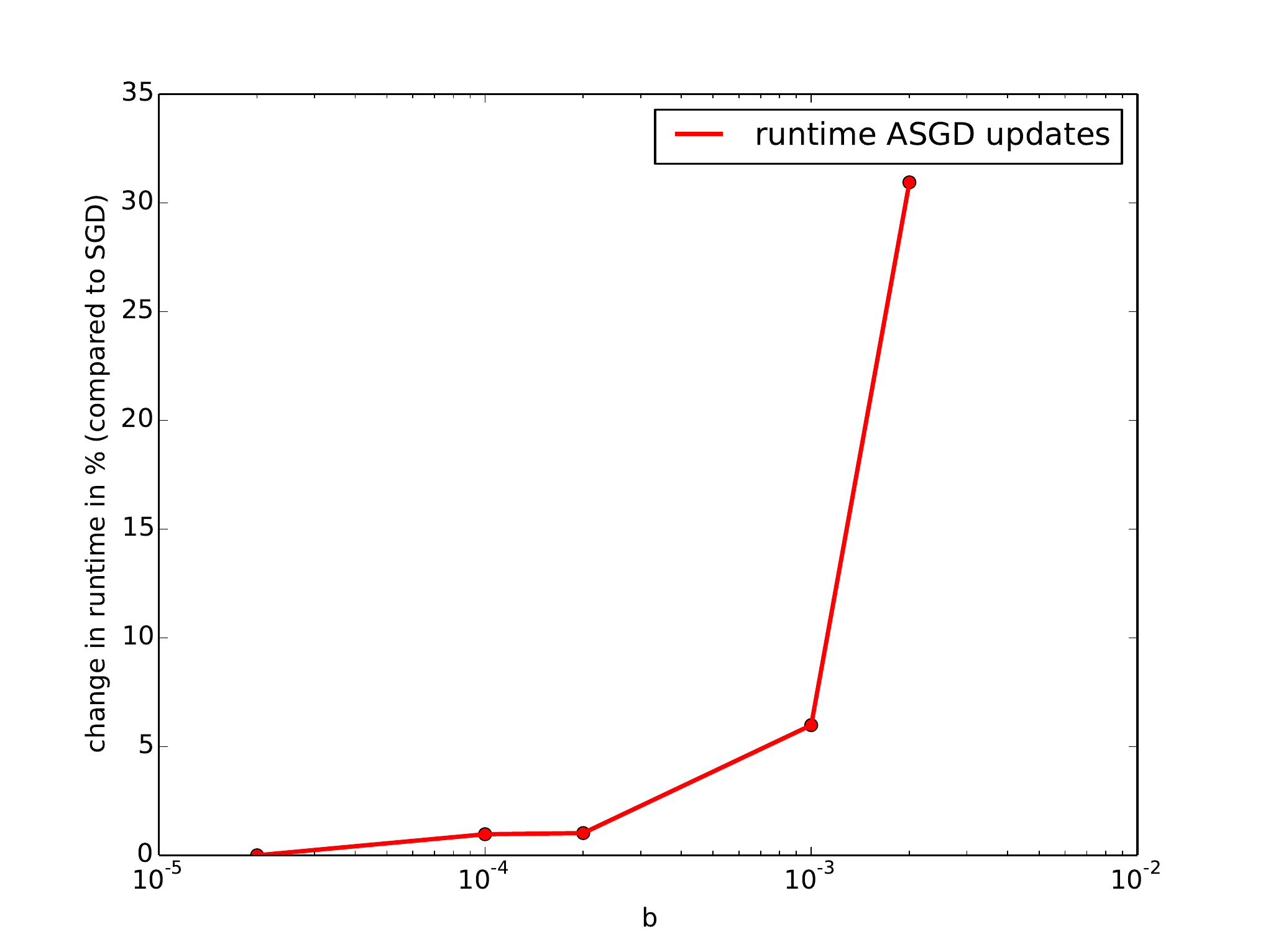}
\caption{Communication cost of ASGD. The cost of higher communication 
frequencies $1\over b$ in ASGD updates compared to communication free SGD updates. 
\label{fig_eval_b}}
\end{figure}

Theoretically, more communication should lead to better results of the ASGD 
algorithm, as long as the node interconnection provides enough bandwidth.
Figure \ref{fig_eval_b} shows this effect: as long as the bandwidth suits the
the communication load, the overhead of an ASGD update is marginal compared to 
the SGD baseline. However, the overhead increases to over $30\%$
when the bandwidth is exceeded.         

\begin{figure}[!ht]
\includegraphics[width=\linewidth]{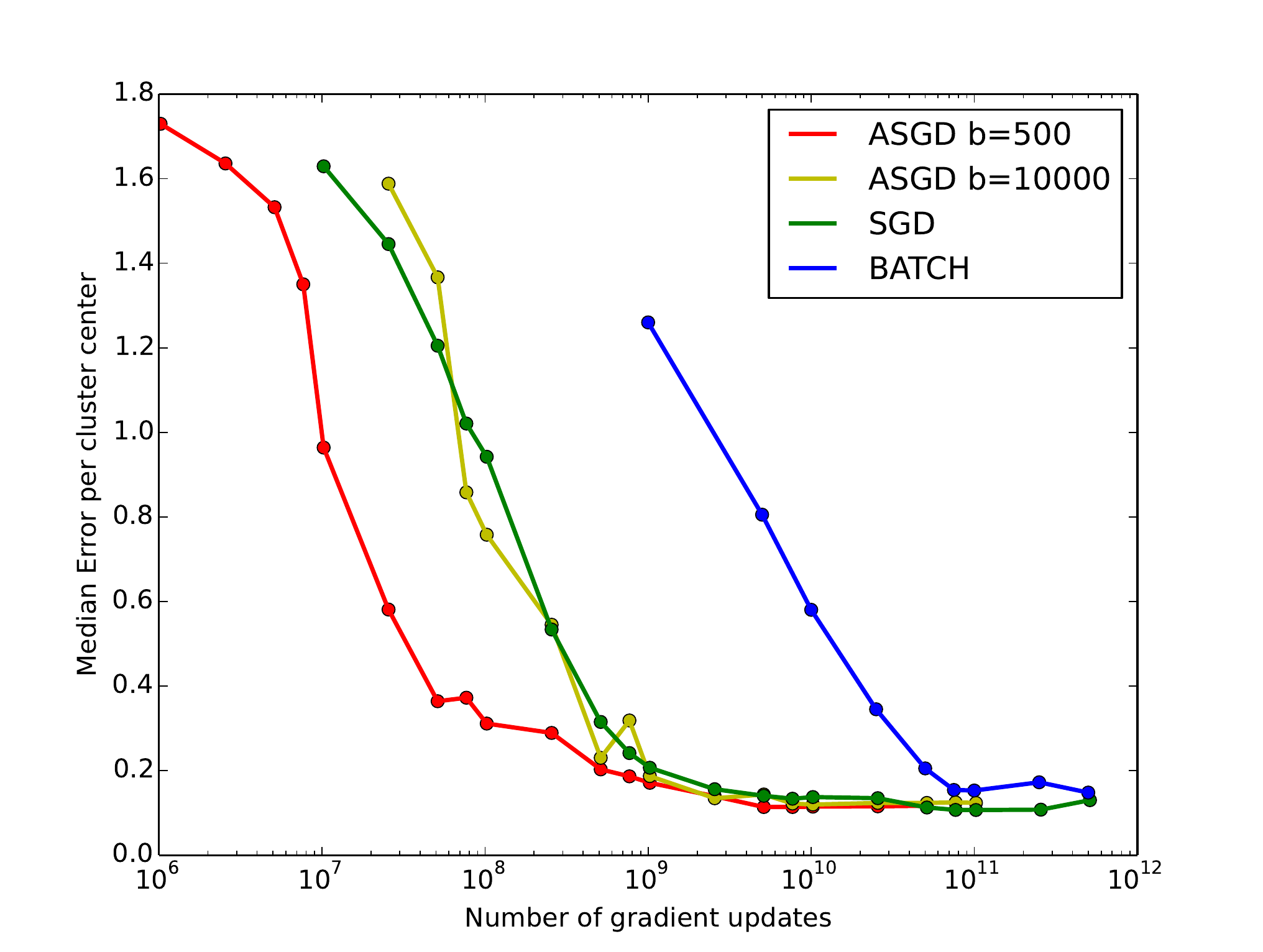}
\caption{Convergence speed of ASGD with a communication frequencies of $1\over 100000$ 
compared to $1\over 500$ in relation to the other methods. Results on Synthetic 
data with $D=10, k=100$.\label{fig_eval_comim}}
\end{figure}

As indicated by the date in figure \ref{fig_eval_b}, we chose $b=500$ for all of our 
experiments. However, as noted in section \ref{sec_balance}, an optimal choice 
for $b$ has to be found for each hardware configuration separately.
\begin{figure}[!ht]
\includegraphics[width=\linewidth]{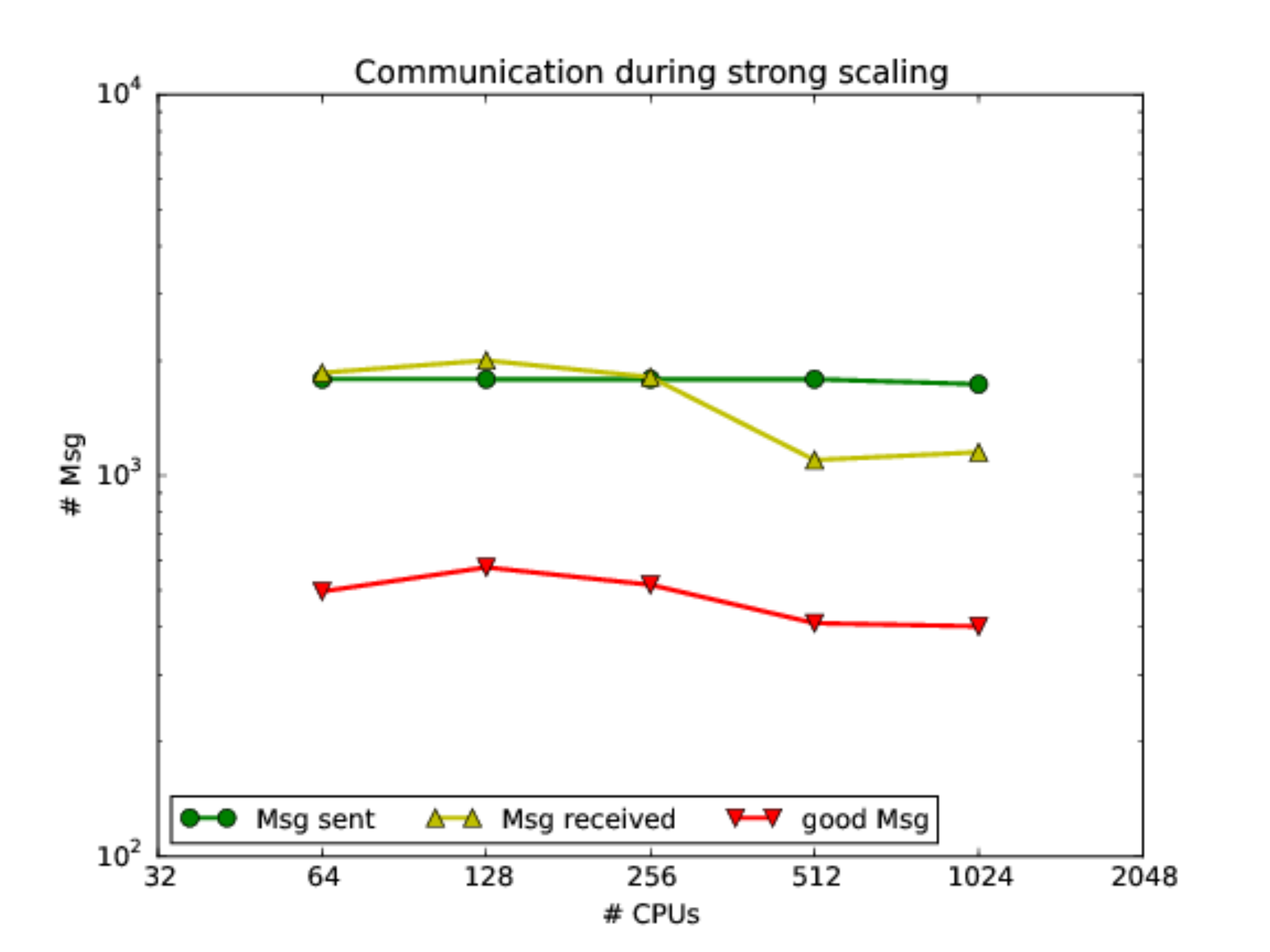}
\caption{Asynchronous communication rates during strong scaling experiment (see
figure \ref{fig_eval_scaling}). This figure shows the average number of messages sent or
received by a single CPU over all iterations. ``Good'' messages are defined as
those, which were selected by the Parzen-window function, contributing to the
local update.
\label{fig_msg}}
\end{figure}
The number of messages exchanged during the strong scaling experiments is shown 
in figure \ref{fig_msg}. While the number of messages sent by each CPU stays 
close to constant, the number of received messages is decreasing. Notably, 
the impact on the asynchronous communication remains stable, because the number of
``good'' messages is also close to constant.    
Figure \ref{fig_eval_comim} shows the impact of the communication frequencies of $1\over b$ on the
convergence properties of ASGD. If the frequency is set to lower values, the convergence 
moves towards the original SimuParallelSGD behavior.

\begin{figure}[!ht]
\includegraphics[width=\linewidth]{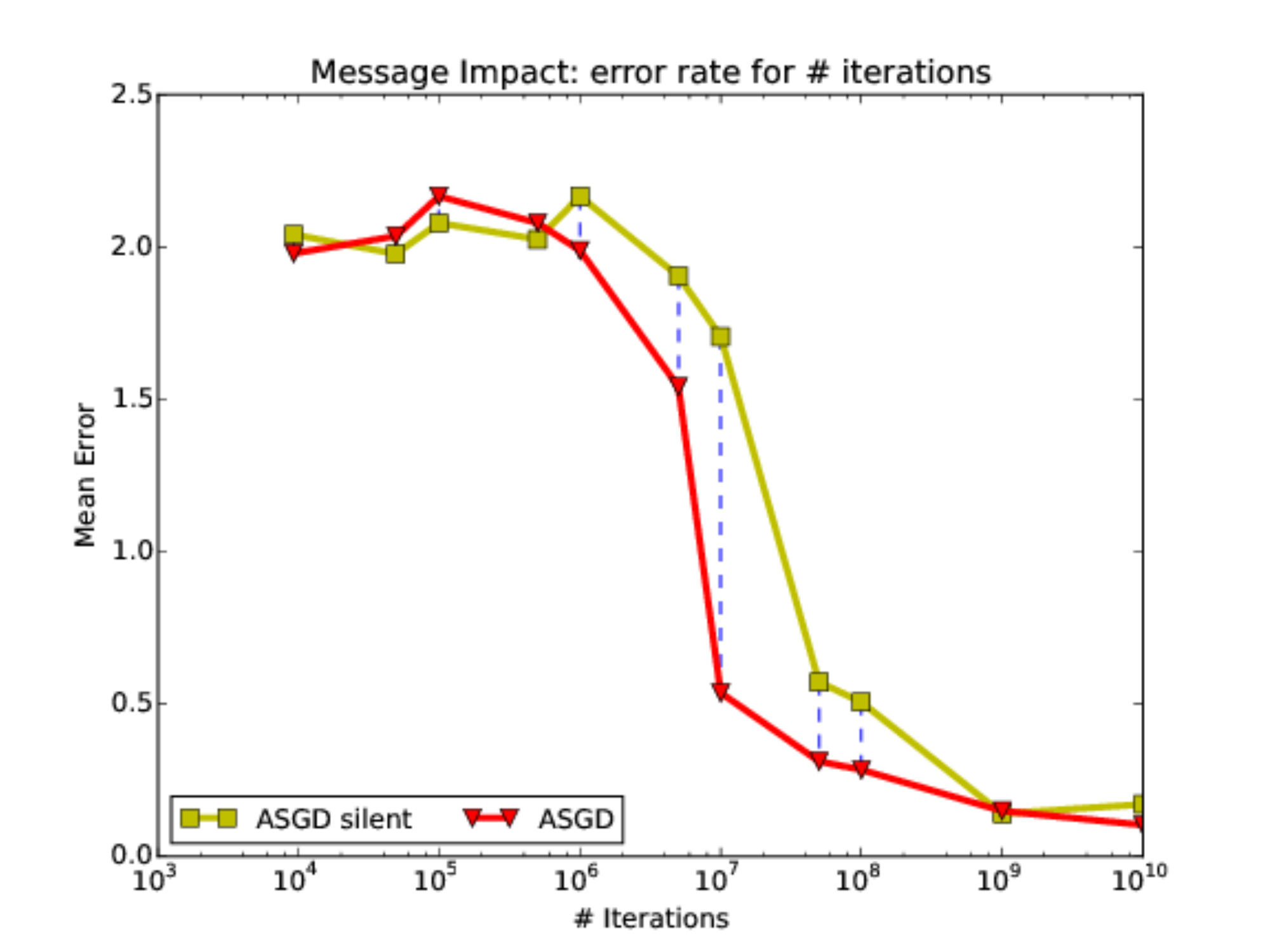}
\caption{Convergence speed of ASGD optimization (synthetic dataset, $k=10,d=10$) 
with and without asynchronous communication (silent). \label{fig_eval_impact_error}}
\end{figure}

\subsubsection*{Impact of the Asynchronous Communication\label{sec_ex_comim}}
\begin{figure}[!ht]
\includegraphics[width=\linewidth]{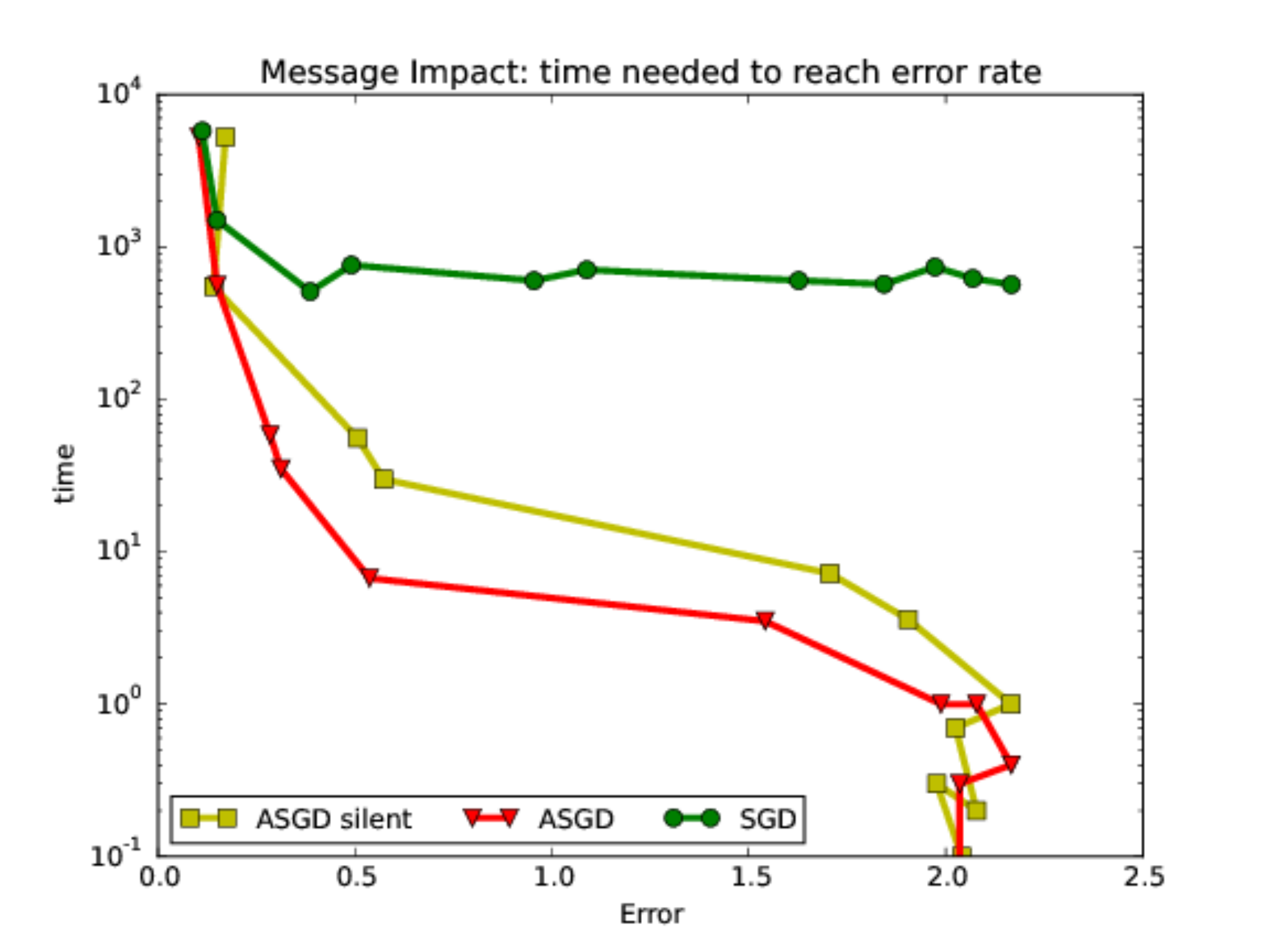}
\caption{Early convergence properties of ASGD without communication (silent) compared
to ASGD and SGD. \label{fig_eval_impact_time}}
\end{figure}

ASGD differs in two major aspects from SimuParallelSGD: asynchronous communication and
mini-batch updates. In order to verify that the single-sided communication
is the dominating factor of ASGD's properties, we simplify turned off the 
communication (silent mode) during the optimization. Figures \ref{fig_eval_impact_error}
and \ref{fig_eval_impact_time} show, that our communication model is indeed 
driving the early convergence feature, both in terms of iterations and time needed
to reach a given error level.

\subsubsection*{Final Aggregation\label{sec_ex_agg}}
\begin{figure}[!ht]
\includegraphics[width=\linewidth]{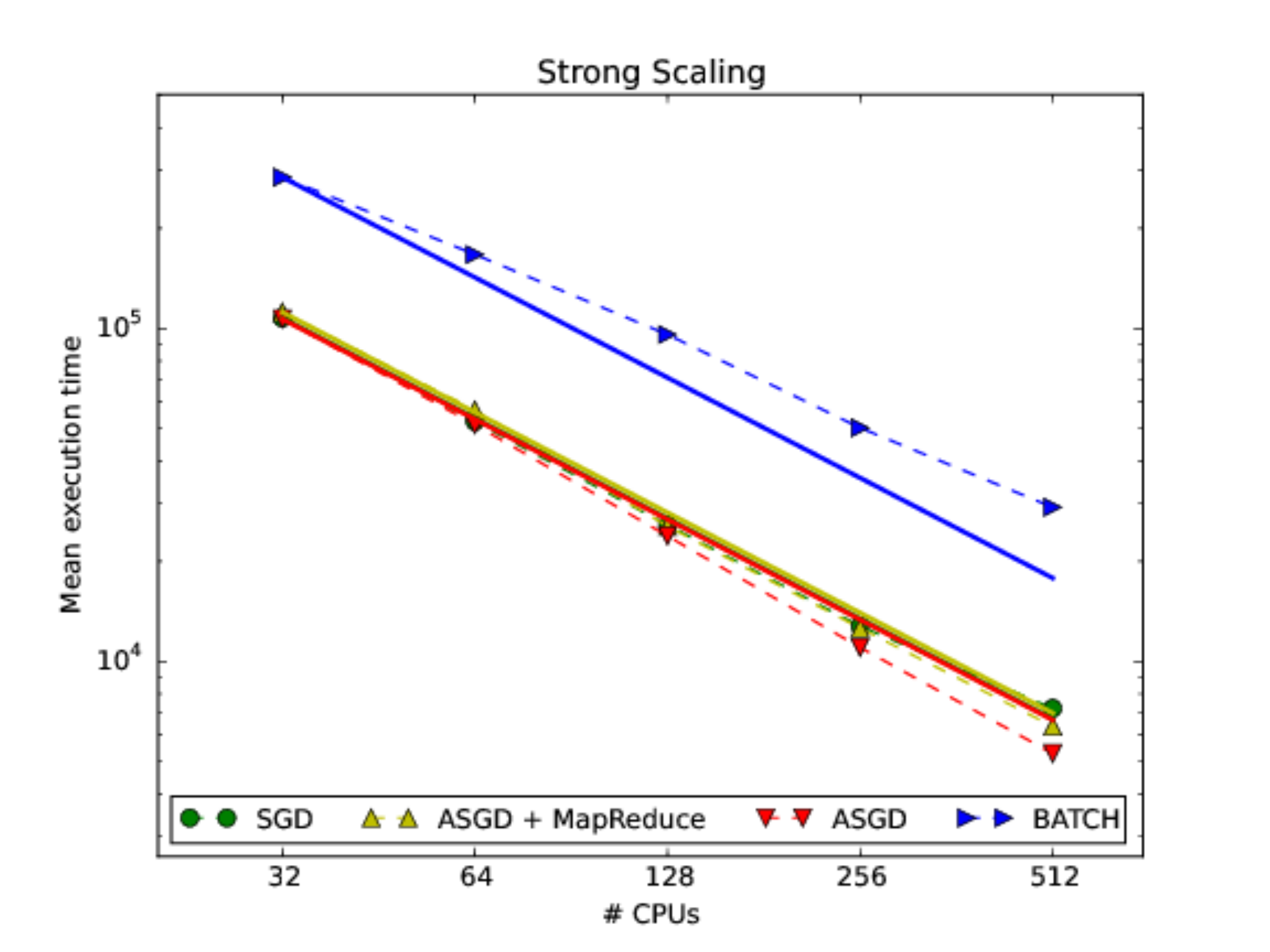}
\caption{Comparison of the runtime and scalability for the 
two possible final aggregation methods of ASGD. 
Synthetic dataset, $k=10,d=10$, and $\sim$1TB data samples. 
Error rates for this experiment
are shown in figure \ref{fig_eval_MR_err}.  
\label{fig_eval_MR_scale}}
\end{figure}

\begin{figure}[!ht]
\includegraphics[width=\linewidth]{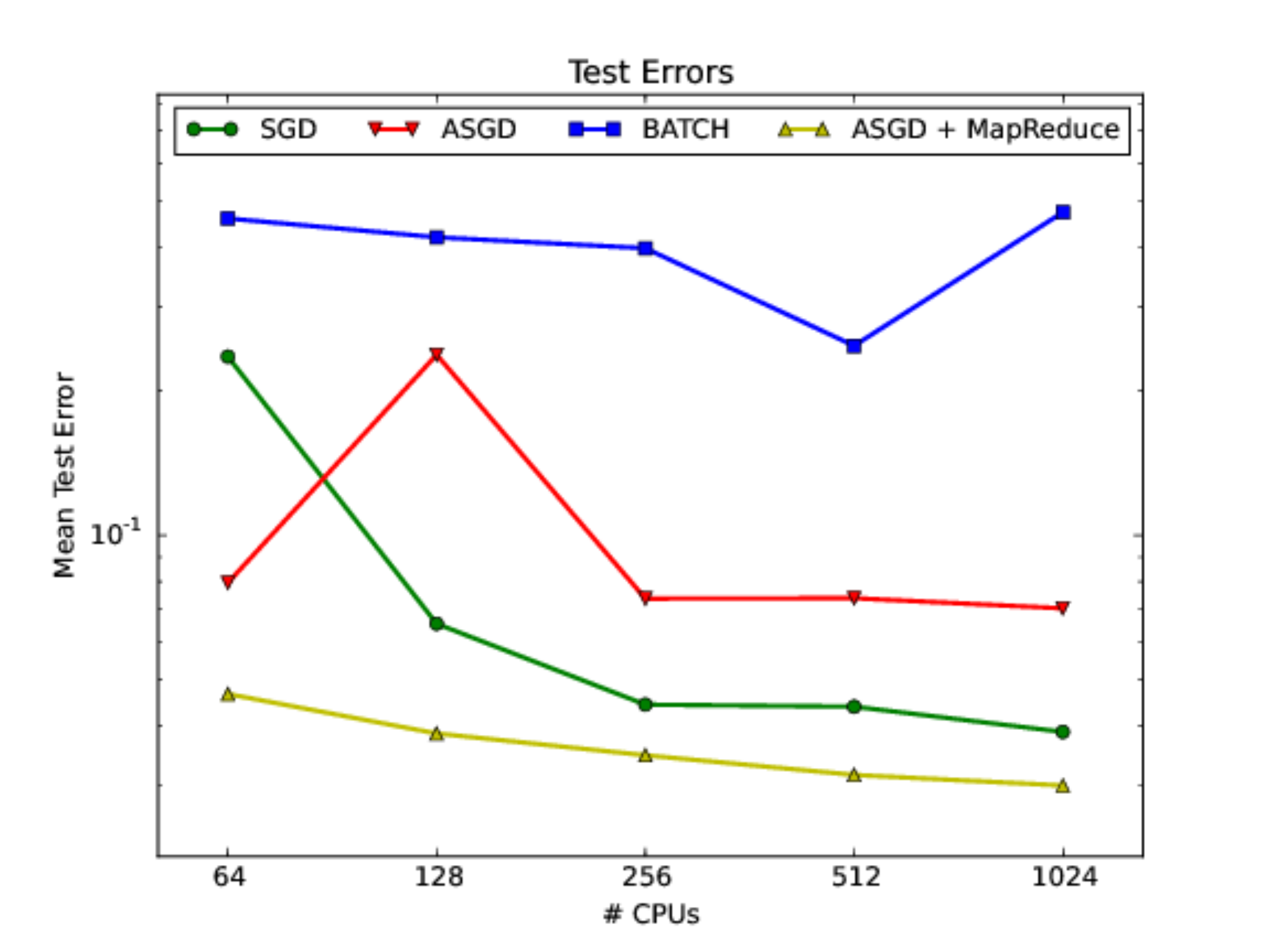}
\caption{Error rates for the same experiment shown in figure \ref{fig_eval_MR_scale}.
Comparing the final aggregation steps of ASGD. 
\label{fig_eval_MR_err}}
\end{figure}

As noted in section \ref{sec_final}, the local results of ASGD could be further 
aggregated by a final reduce step (just like in SGD). Figures \ref{fig_eval_MR_scale}
and \ref{fig_eval_MR_err} show a comparison of both approaches on the
strong scaling experiment (see figure \ref{fig_eval_scaling}).  

\section{Conclusions}
We presented a novel approach towards an effective parallelization of 
stochastic gradient descent optimization on distributed memory systems.
Our experiments show, that the asynchronous communication scheme can 
be applied successfully to SGD optimizations of machine learning algorithms, 
providing superior scalability and convergence compared 
to previous methods.\\
Especially the early convergence property of ASGD should be of high practical
value to many applications in large scale machine learning.  
\bibliographystyle{abbrv}
\bibliography{ASGD}  
\balancecolumns
\end{document}